\documentclass[aps,reprint,twocolumns,pre,superscriptaddress,floatfix,notitlepage,nofootinbib]{revtex4-2}
\usepackage{float}
\makeatletter
\let\newfloat\newfloat@ltx
\makeatother
\usepackage{amssymb,amsmath,amsfonts}
\usepackage{mathtools}
\usepackage{physics}
\usepackage{enumerate}
\usepackage{array}
\usepackage{algorithm}
\usepackage{algpseudocode}
\usepackage{graphicx,epsfig,xcolor}
\usepackage[colorlinks=true,citecolor=blue,linkcolor=red]{hyperref}
\usepackage{newtxtext,newtxmath}
\usepackage{comment}

\begin{document}

\author{\'{A}lvaro Rubio-Garc\'{i}a}
  \email[]{alvaro.rubio@csic.es}
  \affiliation{Instituto de F\'{\i}sica Fundamental, IFF-CSIC, Serrano 113, 28006 Madrid, Spain}
  \affiliation{Inspiration-Q, Calle Tablas de Daimiel, 7,2, 28924, Madrid, Spain}

\author{Samuel Fern{\'a}ndez-Lorenzo}
  \affiliation{Inspiration-Q, Calle Tablas de Daimiel, 7,2, 28924, Madrid, Spain}

\author{Juan Jos{\'e} Garc{\'i}a-Ripoll}
  \affiliation{Instituto de F\'{\i}sica Fundamental, IFF-CSIC, Serrano 113, 28006 Madrid, Spain}

\author{Diego Porras}
  \affiliation{Instituto de F\'{\i}sica Fundamental, IFF-CSIC, Serrano 113, 28006 Madrid, Spain}

\date{\today}

\title{Accurate solution of the Index Tracking problem with a hybrid simulated annealing algorithm}

\begin{abstract}
An actively managed portfolio almost never beats the market in the long term. Thus, many investors often resort to passively managed portfolios whose aim is to follow a certain financial index. The task of building such passive portfolios aiming also to minimize the transaction costs is called Index Tracking (IT), where the goal is to track the index by holding only a small subset of assets in the index. As such, it is an NP-hard problem and becomes unfeasible to solve exactly for indices with more than $100$ assets. In this work, we present a novel hybrid simulated annealing method that can efficiently solve the IT problem for large indices and is flexible enough to adapt to financially relevant constraints. By tracking the S\&P-500 index between the years 2011 and 2018 we show that our algorithm is capable of finding optimal solutions in the in-sample period of past returns and can be tuned to provide optimal returns in the out-of-sample period of future returns. Finally, we focus on the task of holding an IT portfolio during one year and rebalancing the portfolio every month. Here, our hybrid simulated annealing algorithm is capable of producing financially optimal portfolios already for small subsets of assets and using reasonable computational resources, making it an appropriate tool for financial managers. 
\end{abstract}

\maketitle

\section{Introduction}
\label{sec:intro}

It is known that active management of financial portfolios has historically not been able to beat the market consistently in the long term. Together with the high financial costs of active management, this makes that many small investors are now turning their attention into passive management, which focuses on tracking a specific financial index such as the S\&P-500 or NASDAQ and usually have lower transaction fees. To build a passive management portfolio, a reasonable strategy could be to hold all assets inside the index, which would track it in an exact way, but this would also result in very high transaction costs. The task of building this portfolio using only a low number of assets is called the \textit{index tracking} (IT) problem.




The computational difficulty of IT is recognized as a challenge in the literature on this topic, as it is an NP-hard problem\ \cite{Ruiz-Torrubiano2009,Mutunge2018}. There might be external aspects involved in the management of a tracker portfolio, like the quality and treatment of the financial historical data or balancing future beliefs about the market's behavior.
However, regardless of these aspects, the limit on the quality of the resulting tracker portfolio is set by the quality of the algorithm used to solve the IT problem. Indeed, obtaining an exactly optimal solution quickly becomes unfeasible in reasonable time for indices with hundreds of assets.

Because of this hardness, several heuristics have been developed to approximate this combinatorial optimization problem, where the task is both to select a subset of assets to include in the tracker portfolio and also to leverage their weights inside it. 
Let us first mention the broad family of genetic or evolutionary methods: Beasley {\it et al.}\ \cite{Beasley2003} introduced one such algorithm for optimization and tested it on several international financial indexes; Maringer {\it et al.}\ \cite{Maringer2007} used a differential evolutionary algorithm for an empirical study on the Down Jones Industrial Average; and Ruiz-Torrubiano {\it et al.}\ \cite{Ruiz-Torrubiano2009} developed a hybrid algorithm that uses quadratic programming together with a genetic algorithm. 
Another family of methods have addressed the cardinality constraint by selecting assets through a relaxation of the original combinatorial problem. The method by Dose {\it et al.}\ \cite{Dose2005a} groups assets by hierarchical clustering, and assigns weights to the representatives via a much simpler convex optimization problem. The kernel search method by Guastaroba {\it et al.}\ \cite{Guastaroba2012} is a more sophisticated algorithm, where they solve a relaxation of IT with no cardinality constraint and then they use the most relevant weights of the relaxed solution to identify a ``kernel'' of assets, whose actual weights in the portfolio are then found by quadratic programming. Mutunge {\it et al.}\ \cite{Mutunge2018} develop a similar kernel method where assets are introduced one by one in a greedy search.
The class of hybrid methods divides the search process in two parts: a local search over the assets' space and an optimization method to select the weights of the portfolio. Gaspero {\it et al.}\ \cite{Gaspero2011} use a family of greedy local heuristics to select the assets to include in the portfolio. Fern\'andez-Lorenzo {\it et al.}\ \cite{Fernandez-Lorenzo2021} presented a pruning approach in which the selection of a subset of assets is expressed in terms of binary decision variables. Once this selection is made, the weight of each asset in the portfolio is adjusted with quadratic programming.
Recently, Palmer {\it et al.}\ \cite{Palmer2022} explored how the hardness of the IT problem can be tackled with quantum annealing.

The physics-inspired family of annealing methods is a metaheuristic that has enjoyed much success in solving combinatorial optimization problems.
One of such metaheuristics is simulated annealing\ \cite{Kirkpatrick1983} (SA), which is a probabilistic algorithm that travels through the solution space by emulating a physical cooling process in which the system slowly relaxes to a minimum of the cost function. 
SA has actually been applied to a financial task related to index tracking, namely, portfolio selection with a cardinality constraint \cite{Chang2000a,Crama2003}.
Methods based on SA are particularly well suited for optimization with integer variables, and they face some challenges when applied to optimization tasks like the IT problem, where both continuous and binary variables appear.


In this work, we present a novel hybrid method that uses SA in the local search over the assets' space and convex quadratic optimization to select the weights of the portfolio. By using SA to address the combinatorial optimization step of the problem, our algorithm is able to converge in a scalable way into an approximately global optimum,  yielding a quasi-exact numerical solution of the IT problem. We have tested our algorithm by simulations using data from the S\&P-500 index between the years 2011 and 2018 as a benchmark index and we have arrived to the following results:
\begin{itemize}
    \item Our hybrid simulated annealing algorithm is able to find quasi-optimal results for portfolio trackers of sizes between 10 and 30 assets in times that range between 1 second and 20 minutes depending on the size of the portfolio and other market conditions. Our algorithm allows us to consider problems that are intractable with exhaustive solvers such as Gurobi.
    \item We have studied the relationship between the optimized in-sample and out-of-sample tracking results and found that the inherent market noise of out-of-sample results can be lowered by some degree by running large SA computations. 
    \item We have tested our method in a real financial setup by simulating the monthly rebalancing of a portfolio tracker during one year. We have calculated the ex-post tracking error of the tracker portfolio during its active window and found that our algorithm is capable of reaching tracking errors between $2.5-3.5\%$ already with portfolios with 30 assets.
\end{itemize}

Our work is structured as follows. In Sec.\ \ref{sec:index_tracking} we introduce the mathematical definition of the IT problem and discuss the measurement of the tracking error. At the end of the section we also present the treatment of the financial data used in our work. In Sec.\ \ref{sec:sa} we introduce our version of the hybrid SA algorithm, how we tune the algorithm's hyperparameters. We also run a time-to-solution computation and discuss the runtime of hybrid SA. In Sec.\ \ref{sec:lbw} we analyze the optimization of the in-sample tracking error and explore its relation with the out-of-sample tracking error. We also show the average portfolio size needed to obtain a target tracking error. In Sec.\ \ref{sec:rebalancing} we explore the IT problem from a more realistic financial perspective by the tracking error of monthly rebalanced tracker portfolios. Finally, in Sec.\ \ref{sec:conclusions} we lay out the main conclusions of the article.

\section{The IT problem}
\label{sec:index_tracking}

The IT problem consists on selecting a portfolio of $k$ assets from a benchmark index with $L$ available assets ($k<L$), such that the returns of the portfolio follow the returns of the benchmark index as close as possible. We define the portfolio by the weights $\vec{\omega} \in \mathbb{R}^L$ of the assets it holds, which are proportional to the asset's prices at the time when the portfolio was built.

A measurement of the closeness of the tracker portfolio's and index's returns during a specific time window $\mathcal{T}$ is given by the Tracking Error (TE), defined as the standard deviation of the daily difference between the returns of the index and the portfolio 
\begin{equation}
    \begin{split}
            \mathrm{TE}^2(\mathcal{T}) =&\ \textrm{Var}_{t\in\mathcal{T}}\left[r_I(t) - r_p(t)\right] \\
            =&\ \textrm{Var}_{t\in\mathcal{T}}\left[\sum_{i=1}^{L} \left( \omega_i^b - \omega_i \right) r_i (t)\right] \\
            =&\ (\vec{\omega}^b-\vec{\omega})^T \sigma (\vec{\omega}^b-\vec{\omega}),         
    \end{split}
    \label{eq:tracking_error}
\end{equation}
with $r_I(t),r_p(t),r_i(t)$ the returns of the index, the portfolio, and the asset $i$ at time $t$, respectively; $\vec{\omega}^b$ the weights of the benchmark index at the start of the time window $\mathcal{T}$ and $\sigma$ the asset's returns covariance matrix over the time window $\mathcal{T}$. 
Some authors propose to measure the TE as the mean squared error of the difference between the index and tracker returns\ \cite{Beasley2003,Ruiz-Torrubiano2009,Rossbach2011}. Its main point is that a tracker portfolio that has a constant shift in returns with respect to the index would show zero variance. However, as we will see below, we find that for large datasets there is no shift between returns. Other arguments to define the TE using the standard deviation is that it has been shown to produce better out-of-sample portfolios\ \cite{Gnagi2020} and it allows us to work with the covariance matrix $\sigma$ of asset returns\ \cite{Jansen2002,Coleman2006} in order to minimize random noise effects that could potentially spread to the out-of-sample results\ \cite{Laloux1999,Ledoit2003}.

We express the IT problem as a mixed integer quadratic programming (MIQP) problem
\begin{equation}
    \begin{split}
        \textrm{min}&\quad f(\vec{x},\vec{\omega}|\sigma,\Vec{\omega}^b) = (\vec{\omega}-\vec{\omega}^b)^T \sigma (\vec{\omega}-\vec{\omega}^b) \\
        \textrm{s.t.}&\quad \sum_i x_i = k \\
        &\quad \sum_i x_i \omega_i = 1 \\
        &\quad 0 \leq \omega_i \leq x_i
    \end{split}
    \label{eq:miqp_problem}
\end{equation}
with $x_i \in \left\lbrace 0, 1\right\rbrace$ a set of $L$ binary decision variables that are 1 if asset $i$ is included in the portfolio and 0 otherwise. The first condition sets $k$ as the maximum allowed number of assets in the portfolio, while the second and third conditions enforce using the whole budget to build the portfolio and forbid short-selling of assets, respectively. In general, MIQP are NP-hard problems and to solve it we introduce a variant of the hybrid SA algorithm that we explain in the next Section. We note that every other variant of this problem for which the minimization objective is expressed as a convex problem can be solved with out algorithm. An example of which could be the introduction of proportional transaction costs.

In many situations, we want to build a portfolio that tracks a benchmark index in the future. In that case, we assume that, in the absence of market shocks, the asset's returns behave similarly during small time windows. Therefore, we expect that a portfolio that minimizes the TE over past returns (in-sample) will also approximately minimize the TE over a small time window of future returns (out-of-sample), typically a few weeks or months. 

\subsection{Data, covariance matrix and benchmark weights}

In this work we focus on tracking the S\&P-500 financial index between the years 2011 and 2018. Because assets are included and excluded off the index depending on their capitalization, we discard every asset that has not been contained in the index continuously from 2008/01/01 to 2022/02/15, which leaves us with $L=433$ stocks that compose the index. For these stocks we have gathered daily closing price data between these dates from the Yahoo Finance database. Similarly, we have gathered the daily closing price of the S\&P-500 index between these dates (\^\,GSPC ticker).

To sample the benchmark weights $\Vec{\omega}^b$ for a particular time period, we choose a Look Back Window (LBW) of two years and select the weights that minimize the variance of the difference between the prices of a hypothetical portfolio with $\Vec{\omega}^b$ and the index's price $p_I(t)$ during the LBW
\begin{equation}
    \begin{split}
        \textrm{min}&\quad \textrm{Var}_{t\in \textrm{LBW}}\left[p_I(t) - \sum_{i=1}^{L} \omega_i^b p_i (t)\right] \\
        \textrm{s.t.}&\quad \omega_i^b \geq 0,
    \end{split}
    \label{eq:sampling_wb}
\end{equation}
with $p_i(t)$ the price of asset $i$ at time $t$. We minimize the variance of the difference between the prices because we have observed that the portfolios generated with $\vec{\omega}^b$ tend to track the index better using daily prices than using daily returns. To compute the weights we choose a LBW of 2 years in particular because: (a) considering an average of 252 active trade days per year, the number of historical data points to sample is relatively similar in size to the number of stocks considered in the index ($L=433$), which helps to avoid overfitting of the benchmark weights; and (b) the window is sufficiently small such that only recent market trends are considered.

We sample the covariance matrix of the asset returns using again a LBW of 2 years, but in this case we perform an exponential weight averaging of the most recent returns
\begin{equation}
    \sigma_{ij} = \frac{\sum_{t = t_0}^{t_f} \alpha^{t_f-t}\left(r_i(t) - \overline{r}_i\right)\left(r_j(t) - \overline{r}_j\right)}{\sum_{t = t_0}^{t_f} \alpha^{t_f-t}},
    \label{eq:cov_matrix}
\end{equation}
with $t_0, t_f$ the time boundaries of the LBW, $\overline{r}_i$ the mean value of the returns of asset $i$ over the LBW without exponential averaging, and $\alpha$ a constant that we define in terms of a half-life $\tau$ of the exponential weights, $\alpha = 2^{-1/\tau}$. By shifting the parameter $\tau$ we can effectively reduce the size of the LBW from two years to several weeks, which can put the focus of the IT problem on tracking only the most recent market trends. We note that the limit $\tau \rightarrow \infty$ corresponds to the usual sample covariance matrix.

\section{Hybrid Simulated Annealing}
\label{sec:sa}

Simulated annealing\ \cite{Kirkpatrick1983} is an algorithm widely used in combinatorial optimization problems. It is a variant of the Metropolis-Hastings algorithm in which the temperature of the target distribution, usually a Boltzmann distribution, is lowered smoothly until the system remains frozen in different local minima. For combinatorial optimization problems there exist proofs that guarantee its convergence to the global minimum for an asymptotic number of Metropolis steps and particular temperature schedules\ \cite{Mitra1986}.

The standard SA algorithm is difficult to implement for MIQP problems, as its domain consists of a discrete space and a continuous space. In this work we present a hybrid variation of the usual SA algorithm that is targeted to solve MIQP problems with cardinality constraints. We start by noting that if we fix the discrete variables $\Vec{x}$, then the task of minimizing the continuous variables $\Vec{\omega}$ becomes a quadratic programming (QP) problem that can be solved efficiently in polynomial time with state-of-the-art solvers. This is because the covariance matrix $\sigma$ is positive semidefinite. Thus, we propose a two-step hybrid SA algorithm where the discrete variables $\vec{x}$ are optimized using SA and then the continuous variables $\vec{\omega}$ are optimized using a QP problem solver.

We provide a scheme of our algorithm in Alg.\ \ref{alg:sa_alg}. We start with an initial portfolio with $k$ random assets. At each step $s$, we draw from a uniform distribution an asset $a$ outside the portfolio and an asset $b$ inside the portfolio. Then, we propose a new portfolio $\vec{x}^\prime$ with asset $a$ inside the portfolio and asset $b$ outside (lines 6-8). This keeps the cardinality fixed, $\sum_i x_i = \sum_i x_i^\prime$. Then we compute the cost function of the new proposed configuration $\Vec{x}^\prime$ by solving a QP problem that optimizes the weights $\vec{\omega}^\prime$ of the new portfolio (lines 10-11). Finally, we accept the proposed configuration $\vec{x}^\prime,\vec{\omega}^\prime$ using the Metropolis-Hastings' acceptance rule by comparing the difference between the proposed and old cost functions, where the acceptance probability depends on the annealing temperature at that step $\beta_s$ (lines 13-18). The algorithm stops when $N$ steps have been computed. Due to the stochastic nature of the algorithm, we run $n$ independent copies. We scripted the algorithm in a first version using the Julia language and used the open-source package COSMO.jl\ \cite{Garstka2021} to solve the QP problem, and entirely in Cython in the second version, which we used to write a fast QP solver.

\begin{algorithm}[b]
\caption{Hybrid Simulated Annealing for IT}\label{alg:sa_alg}
\begin{algorithmic}[1]
\Procedure {HybridSimulatedAnnealing}{$\vec{x},\sigma,\Vec{\omega}^b, k, N, \Vec{\beta}$}
    \State $\vec{\omega} \gets \textrm{argmin}_{\vec{\omega}} f(\vec{\omega}|\vec{x},\sigma,\Vec{\omega}^b),\ \textrm{s.t.} \sum_i x_i\omega_i = 1,\ \Vec{\omega}\succeq 0$
    \State $C\left(\Vec{x}\right) \gets f(\Vec{x},\Vec{\omega})$ 
    \For{$s \gets 1,N$}
        \State \Comment{Propose new $\vec{x}$}
        \State $a \gets \textrm{Uniform}(r=1,2,\dots,L\,|\,x_r=0)$
        \State $b \gets \textrm{Uniform}(r=1,2,\dots,L\,|\,x_r=1)$
        \State $\vec{x}^\prime \gets \vec{x},\ x_a^\prime \gets 1,\ x_b^\prime \gets 0$
        \State \Comment{Evaluate QP problem}
        \State $\vec{\omega}^\prime \gets \textrm{argmin}_{\vec{\omega}} f(\vec{\omega}|\vec{x}^\prime,\sigma,\Vec{\omega}^b),\ \textrm{s.t.} \sum_i x_i^\prime\omega_i = 1,\ \Vec{\omega}\succeq 0$
        \State $C\left(\Vec{x}^\prime\right) \gets f(\Vec{x}^\prime,\Vec{\omega}^\prime)$
        \State \Comment{Acceptance step}
        \State $p \gets \textrm{Uniform}[0,1)$
        \If {$p \leq \min \left\lbrace 1,\ e^{-\beta_s \left(C\left[\Vec{x}^\prime\right) - C\left(\Vec{x}\right)\right]} \right\rbrace$}
            \State $\vec{x} \gets \vec{x}^\prime,\ \vec{\omega} \gets \vec{\omega}^\prime$ \Comment{Accept new portfolio}
        \Else
            \State $\vec{x} \gets \vec{x},\ \vec{\omega} \gets \vec{\omega}$ \Comment{Reverse to previous portfolio}
        \EndIf
    \EndFor
    \State \textbf{return} $\Vec{x}$, $\vec{\omega}$
\EndProcedure
\end{algorithmic}
\end{algorithm}

\begin{figure*}[t]
    \centering
    \includegraphics[width=\linewidth]{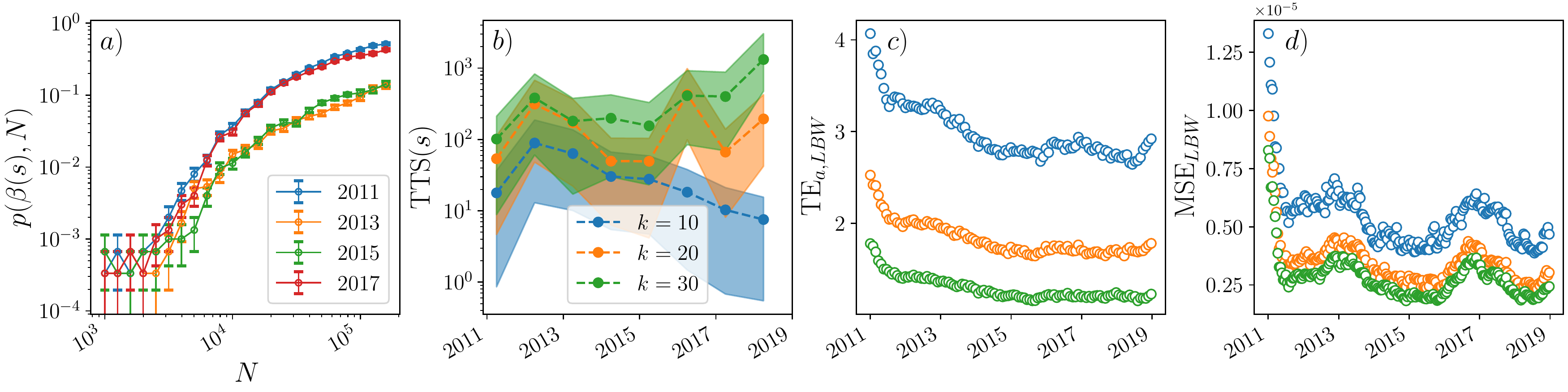}
    \caption{(a) Probability $p(\beta(s),N)$ of finding the optimal TE using portfolios with basket size $k=10$ for different years against the number of steps $N$. (b) TTS estimation for solving the IT problem for the S\&P-500 index using hybrid SA for different basket sizes $k$. The shaded areas represent a $95\%$ confidence interval. (c) Annualized TE over a LBW of 2 years with the same basket sizes $k$ as the previous panel. (d) Mean squared error of the difference between the index' and portfolio's returns over the LBW.}
    \label{fig:fig1}
\end{figure*}

\subsection{Hyperparameter tuning}

The choice of hyperparameters is crucial for hybrid SA to be able to produce high quality portfolios. These are: the choice of the temperature schedule $\beta(s)$, the number of steps $N$ and the number of independent copies $n$. We perform the hyperparameter tuning as follows. First, we select multiple choices of initial and final temperatures $\beta(1),\beta(N)$ and how to go from $\beta(1)$ to $\beta(N)$: using a linear, inverse, logarithmic or exponential form. After that, we run a hybrid SA simulation for every schedule with a high number of steps $N\sim 10^5-10^6$ and copies $n\sim 100$. Then, we compute for each schedule the ratio of solutions that reached the optimal TE. Finally, we fix the temperature schedule as the one for which the ratio of optimal TE found was highest and run a Time-To-Solution (TTS) computation\ \cite{Ronnow2014} with this schedule to determine the optimal choice of steps $N$ and copies $n$.

A TTS computation starts by treating a single hybrid SA simulation with hyperparameters $\left(\beta(s),N\right)$ as a Bernoulli process with a probability $p(\beta(s),N)$ of finding an optimal TE. If a single SA run yields the ground state with probability $p(\beta(s),N)$, then the number of repetitions $R(\beta(s),N;P)$ one needs to find the optimal TE with a probability $P$ is given by
\begin{equation}
    R(\beta(s),N;P) = \frac{\log(1-P)}{\log[1-p(\beta(s),N)]}.
\end{equation}
We set $P=99\%$ for the rest of this work. The total computation time $T(\beta(s),N;P)$ needed to output an optimal TE with probability $P$ is thus given by the runtime of a single hybrid SA times the number of repetitions $T(\beta(s),N;P) \propto N\cdot R(\beta(s),N;P)$. The TTS is then defined as the minimum total computation time
\begin{equation}
    \textrm{TTS} = \min_{\beta(s),N} \left\lbrace T(\beta(s),N;P=99\%) \right\rbrace.
\end{equation}

In the context of our work, the number of available assets $L=433$ makes it unfeasible to obtain the global minimum TE for the number of shares $10 \leq k \leq 30$ considered in our tracker portfolios. Thus, we define the optimal TE as the best TE found for the largest number of steps, typically $N \sim 10^5$ for $k=10$ and $N \sim 10^6$ for $k=20,30$. While we cannot guarantee to have found the optimal TE, we have checked that our hybrid SA algorithm is capable of finding the optimal TE for smaller indices of size $L \leq 100$ and comparing the TE found with an exhaustive solver such as Gurobi.

We show in Figs.\ \ref{fig:fig1}(a, b) the TTS estimation of a set of IT computations using the S\&P-500 index as a benchmark between the years 2011 and 2018. For each computation, we sample the covariance matrix $\sigma$ over a LBW $\mathcal{T}$ of 2 years with a half-life of $\tau = 2y$. For this data, we have found that hybrid SA yields optimal results with the following exponential temperature schedule
\begin{equation}
    \log_{10}\left(\beta(s)\right) = \frac{13 + k}{20} + \frac{3}{20}\frac{s-1}{N-1}.
\end{equation}
In Fig.\ \ref{fig:fig1}(a) we show the probability $p(\beta(s),N)$ of finding the optimal TE for a basket size of $k=10$ assets for different years (colors). We observe that already for $N\sim 10^4$ steps there is a $\sim1\%$ chance of finding an optimal portfolio, while the probability increases at a relatively slow pace after that, and more or less saturates after $N\sim10^5$ steps. We show in Fig.\ \ref{fig:fig1}(b) the resulting TTS estimation in seconds for different years and basket sizes of $k=10,20,30$ assets (colors). The shaded areas represent a $95\%$ confidence interval. We observe that the TTS can range from 1 second to 20 minutes, with the TTS increasing for large $k$. These runtimes make it accessible to solve the IT problem in small workstations and thus, our hybrid SA algorithm can be easily used by financial managers. There is also the possibility of speeding up computations by using parallelization of the independent $n$ runs of the algorithm and also suboptimal solutions can be obtained already for smaller number of steps $N$ than optimal, as we observe in the slow convergence in Fig.\ \ref{fig:fig1}(a).

\begin{figure*}[t]
    \centering
    \includegraphics[width=\linewidth]{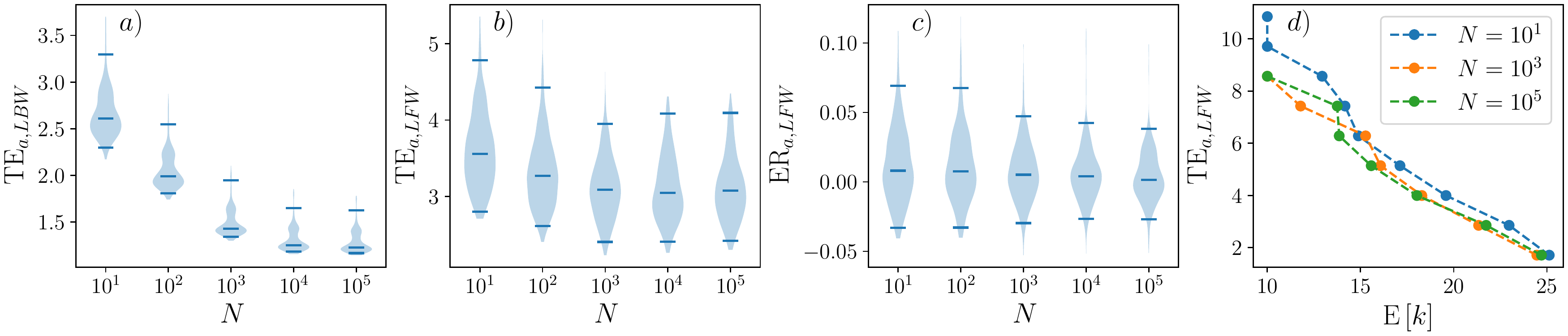}
    \caption{(a) TE in a LBW of two years for tracking the S\&P-500 index between the years 2011 and 2018 against the number of steps $N$ of the hybrid SA simulations. Here $\tau=2y$ and $k=30$. (b, c) TE and ER in a LFW of 1 year against the number of steps $N$ of SA. (d) TE in the LFW against the mean portfolio size $\mathrm{E}\left[k\right]$ for different $N$.}
    \label{fig:fig2}
\end{figure*}

\section{In-sample and out-of-sample Tracking Error}
\label{sec:lbw}

In this Section we show the results of solving the IT problem with the previously found optimal hyperparameters to track the S\&P-500 index between the years 2011 and 2018. We use the benchmark weights and covariance matrices that have been described at the end of Sec. \ref{sec:index_tracking}, where we use a half-life $\tau$ of the covariance matrix of two years. We choose this $\tau$ in particular because the effective size of the LBW window is of 2 years and it provides the best results when doing rebalancing experiments, as we observe in the next section.

We show in Fig.\ \ref{fig:fig1}(c) the result of optimizing the TE in-sample over a LBW of 2 years for different basket sizes $k=10, 20, 30$. We present our results in terms of the annualized TE\ \cite{Dose2005a}
\begin{equation}
    \textrm{TE}_{a} = 100\cdot \textrm{TE} \sqrt{252}, 
    \label{eq:rescaled_te}
\end{equation}
which corresponds to the total TE that the portfolio would accumulate in one year (252 market days) given in percentage points. Each point corresponds to a LBW that ends at the last active day of one week of its corresponding year. We observe that the TE decreases as the basket size $k$ increases, as expected, because having more assets can make the tracker portfolio more representative of the index. We also observe that for sizes $k=20,30$, the TE lays between $2\%$ and $3\%$, which can represent an acceptable TE for a financial manager. 

As argued in Refs.\ \cite{Beasley2003,Ruiz-Torrubiano2009,Rossbach2011}, a portfolio that minimizes the tracking error variance could still show a constant shift in the evolution of the returns with respect to the original index. This shift can be represented by the mean squared error between the returns of the index and the portfolio
\begin{equation}
   \mathrm{MSE}_{\mathcal{T}} = \frac{1}{n_d} \sum_{t\in \mathcal{T}} \left[ r_I(t) - r_p (t)\right]^2.
    \label{eq:mse}
\end{equation}
We plot this error in Fig.\ \ref{fig:fig1}(d) for the same simulations as in Fig.\ \ref{fig:fig1}(c). Our results do not show any relevant constant shift between the returns of the index and the portfolio, which justifies the use of the standard deviation as the definition of the TE.


\subsection{Out-of-sample results and optimization strength}

There is always an inherent stochastic noise in market behaviors, which can result in portfolios that closely track an index in-sample but do not work well out-of-sample. 
To check to what extent an accurate solution of the MIQP problem in-sample is relevant for the out-of-sample performance of the tracker portfolio, we have repeated the above computations with different numbers of annealing steps $N$ to observe its effect on the out-of-sample results. For these computations we keep the number of copies fixed, $n=150$.

First, we show in Fig.\ \ref{fig:fig2}(a) the in-sample annualized TE over a LBW of 2 years between 2011 and 2018 for different optimization strengths, $N=10,100,\dots,10^5$, for $k=30$. The latter number of steps $N=10^5$ is close to the optimal estimated by our TTS computations and represent full in-sample optimization. For each optimization strength, all data points have been represented as a violin plot (with roughly 400 samples per violin plot), of which the middle bar represents the median, the outer bars represent the 95\% confidence interval and the shaded area represents the density of TE points. As we expected, increasing the optimization strength decreases the median in-sample TE until it starts to saturate after $N=10^4$ steps.

\begin{figure*}[t]
    \centering
    \includegraphics[width=\linewidth]{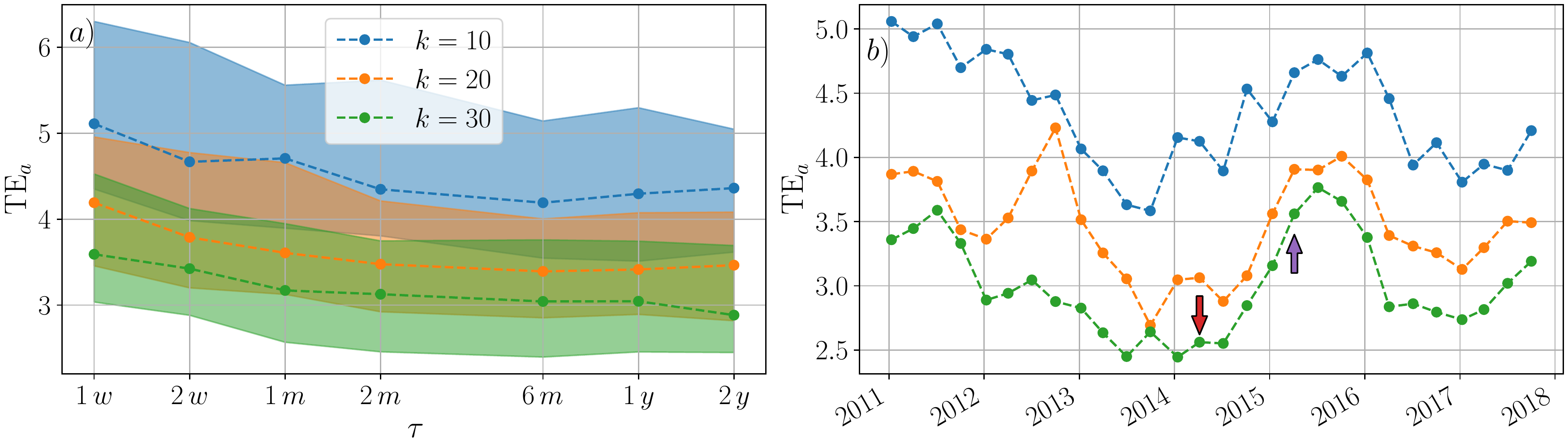}
    \caption{(a) Median annualized TE and 95\% confidence interval (shaded area) of rebalanced portfolios between the years 2011 and 2018 for different basket sizes $k$ and using different values of $\tau$ to compute the covariance matrix. The portfolios are hold during one year and are rebalanced every month. (b) Annualized TE of rebalanced portfolios with a $\tau$ of two years for different $k$. The red and purple arrows indicate the TE of the two portfolios shown in Fig.\ \ref{fig:fig4}.}
    \label{fig:fig3}
\end{figure*}

In Fig.\ \ref{fig:fig2}(b) we show the out-of-sample annualized TE if we hold the tracker portfolio for a LFW of 1 year. Here, the improvement with the number of steps is not as drastic as with in-sample TE. However, we still observe an improvement of the out-of-sample TE when we increase the optimization strength, which seems to saturate after $N=10^3$ steps. After this point, market noise seems to spoil any gains in the optimization of hybrid SA.

An alternative metric to measure the quality of a tracker portfolio is the annualized Excess Return (ER), which measures the difference between the index's and the portfolio's annualized cumulative returns
\begin{equation}
    \mathrm{ER}_{a}(n_d) = \left(r_p\right)^{252/n_d} - \left(r_I\right)^{252/n_d},
\end{equation}
with $r_p, r_I$ the cumulative returns of the portfolio and the index, respectively, at the end of a time window given by its number of days $n_d$. In this case, a perfect tracker portfolio should have zero ER with the benchmark index. We plot in Fig.\ \ref{fig:fig2}(c) the annualized out-of-sample ER of holding the tracker portfolio during 1 year against the optimization strength. We observe that a greater optimization strength results in smaller ER and that there is a small gain in computing $N=10^5$ steps over $N=10^3$. Together with the out-of-sample TE results of Fig.\ \ref{fig:fig2}(b), this indicates that a good optimization of the in-sample TE will result in tracker portfolios of good quality, lowering (although not vanishing) the effects of market noise.

Another observable of financial interest is the mean basket size necessary to reach a threshold out-of-sample TE, which is directly related to the costs of purchasing a particular portfolio. We show in Fig.\ \ref{fig:fig2}(d) the mean size $\textrm{E}\left[k\right]$ that a tracker portfolio needs to have to reach a particular annualized TE over one year for different optimization strengths. 
Here, we also computed portfolios with $k=15$ and $25$ assets to get smooth results. 
We observe that increasing the accuracy of the SA optimization algorithm actually results in a reduction of the size (by 2-3 assets) of the tracking portfolios needed for a target annualized TE.
While this effect seems to saturate after $N=10^3$ steps, this indicates that we can build cheaper tracker portfolios with the same TE quality by just increasing the optimization strength of hybrid SA.

\section{Portfolio rebalancing using hybrid SA}
\label{sec:rebalancing}

In general, the composition of a tracker portfolio is rebalanced at periodic intervals to adapt it to the newest market trends. In this section, we will simulate the process of holding a tracker portfolio over one year with monthly rebalancing, which is standard in many financial scenarios. To perform the rebalancing, we compute an optimized hybrid SA simulation at the day before which we want to rebalance. The starting date of the analyzed tracker portfolios are the first active day of the $1^{th},\,13^{th},\,26^{th}$ and $39^{th}$ weeks of each year between 2011 and 2018. The considered LBW is 2 years and we set the half-life $\tau$ of the covariance matrix as an optimization parameter to study the optimal effective size of the LBW.

We show in Fig.\ \ref{fig:fig3}(a) the median annualized TE for different basket sizes $k$ and using different $\tau$ values to compute the covariance matrix, from one week (with an effective size of 1 month of the LBW) to two years (with an effective size of 2 years of the LBW). In the figure the shaded areas represent the 95\% confidence interval of the annualized TE. We observe that the TE decreases as $\tau$ becomes larger until it saturates after $\tau=6m$. This implies that in rebalancing scenarios and, in general, in the IT problem, having a large size of the LBW helps to replicate all relevant market trends that will play a role in the future. For $\tau=2y$, the annualized TE lies between $3.5-5\%$ for $k=10$, between $3-4\%$ for $k=20$ and between $2.5-3.5\%$ for $k=30$, which is an acceptable TE range in financial standards. The large variability in the TE is due to some periods being harder to track than others. This is shown in Fig.\ \ref{fig:fig3}(b) where we plot the annualized TE for rebalanced portfolios with  using $\tau=2y$ for the covariance matrix. We observe that the rebalanced portfolios with starting dates in the year, e.g. 2015, with $k=30$ assets result in larger annualized TE than in other years with the same number of assets. This effect has already been observed in Ref.\ \cite{Gnagi2020} and can be linked to market shocks happening during the rebalancing period, which can significantly change the previous trends happening in the market.

\begin{figure*}[t]
    \centering
    \includegraphics[width=\linewidth]{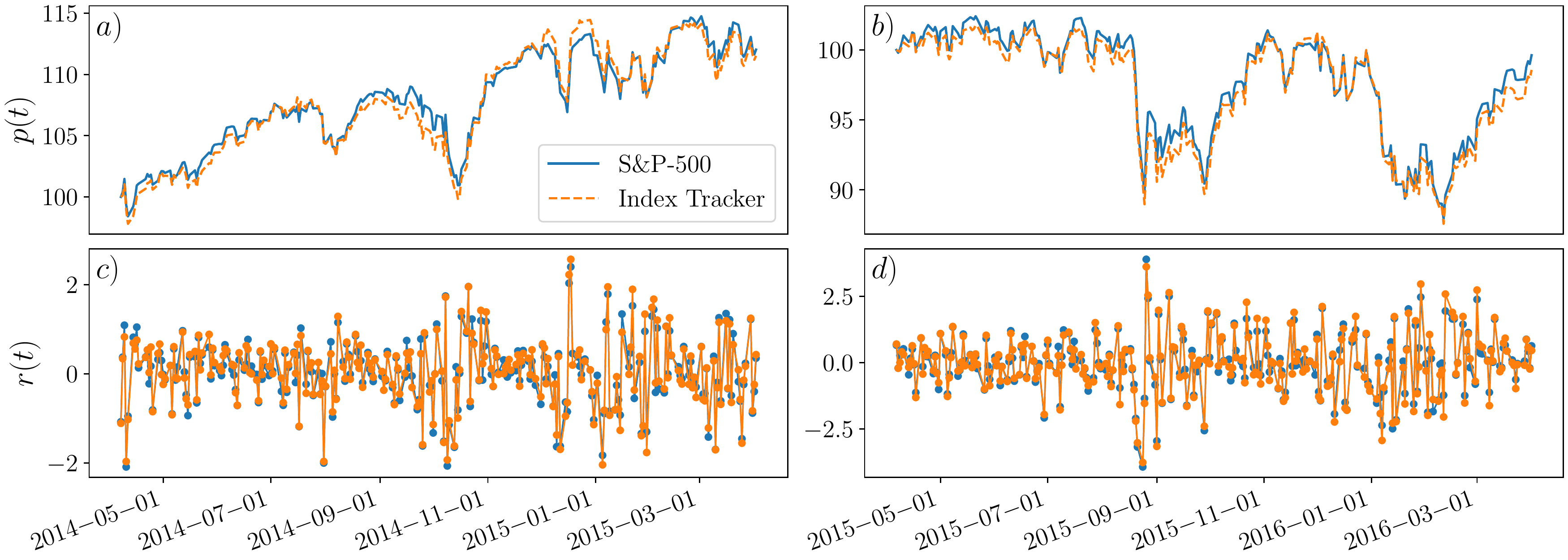}
    \caption{Normalized prices (a) and daily returns (c) of the S\&P-500 index (blue) and a monthly rebalanced portfolio (orange) with $k=30$ assets over 1 year. The holding period startas on $2014/04/06$ and the half-life used to compute the covariance matrix is $\tau=2$ years. Normalized prices (b) and daily returns (d) for a portfolio of the same characteristics with a holding period starting on $2015/04/06$.}
    \label{fig:fig4}
\end{figure*}

We show in Fig.\ \ref{fig:fig4} the normalized daily cumulative returns $p(t)$ and daily returns $r(t)$ of two examples of rebalanced portfolios with $k=30$ assets. In Figs.\ \ref{fig:fig4}(a, c) we show one in a period with no market shocks starting on $2014/04/07$ (red arrow in Fig.\ \ref{fig:fig3}(b)); and in Figs.\ \ref{fig:fig4}(b, d) one starting on $2015/04/06$, which has to deal the $2015/08/24$ flash crash and a $10\%$ market drop at the start of 2016 (purple arrow in Fig.\ \ref{fig:fig3}(b)). We observe how in both cases our tracking portfolio is able to track the S\&P-500 index even in the case of large market downfalls.

\section{Conclusions}
\label{sec:conclusions}

The main result of this work is the introduction of a hybrid Simulated Annealing algorithm that is capable of solving Mixed Integer Quadratic Programming problems with cardinality constraints and is flexible enough to be adapted to general Mixed Integer problems. We applied this algorithm to solve the Index Tracking problem, which falls into the category of NP-hard mathematical problems. In particular, we used it to track the S\&P-500 index between the years 2011 and 2018 using a subset of $L=433$ stocks inside the index and allowing for different numbers of stocks $10 \leq k \leq 30$ to be present in the tracker portfolio. Our hybrid algorithm is capable of finding approximately optimal solutions in the range between one second and 20 minutes of computational runtime and we believe it can work seamlessly with lager indices with thousands of stocks. Our algorithm is thus capable of solving Index Tracking problems for indices with more than hundreds of assets, for which exact solvers like Gurobi would require unfeasible amounts of time.

Using our algorithm we have studied the relation between minimizing the Tracking Error in-sample and the resulting Tracking Error out-of-sample. We have found that there is an advantage in making big simulations with large number of steps and copies, although the market noise makes that some observables of the out-of-sample portfolios start to saturate after medium step sizes $N=10^3$. While this is unavoidable, we have found that increasing the optimization strength can also have a positive effect on the size of the tracker portfolio, effectively reducing the mean portfolio size needed to reach a target Tracking Error. Such noise effects could be potentially avoided in future work by computing refined in-sample covariance matrices that lower the stochastic noise of past returns.

Finally, we performed a series of observations that are related to how a financial manager would potentially manage an Index Tracking portfolio with monthly rebalancing. We computed several tracker portfolios that were held over one year and found that portfolios with $k=30$ assets can already result in an annualized Tracking Error in the range of $2.5-3.5\%$, which can be regarded as a good objective in many financial applications. We stress that providing more refined covariance matrices to the hybrid Simulated Annealing algorithm could result in even better tracker portfolios. We believe this makes our algorithm very suitable for its use in financial environments.

\section{Acknowledgments}
We acknowledge the CSIC Interdisciplinary Thematic Platform (PTI+) on Quantum Technologies (PTI-QTEP+), and 
project PID2021-127968NB-I00 funded by MCIN/AEI/10.13039/501100011033/FEDER,UE. 
This
research is part of the CSIC program for the Spanish Recovery, Transformation and Resilience Plan funded by the Recovery and Resilience Facility of the European Union, established by the Regulation (EU) 2020/2094. 
The authors also gratefully
acknowledge the Scientific computing Area
(AIC), SGAI-CSIC, for their assistance while using the DRAGO Supercomputer for performing
the simulations, and Centro de Supercomputación de Galicia (CESGA) who provided
access to the supercomputer FinisTerrae.
\bibliography{Index-tracking}

\end{document}